# Giant Magnetoresistance Effect in Organic Material and Its Potential for Magnetic Sensor

Mitra Djamal[1], Ramli[2], Sparisoma Viridi[3] and Khairurrijal[4]
[1]Department of Physics, Institut Teknologi Bandung, Bandung, Indonesia
(Tel : +62-821-1659-1960; E-mail: mitra@fi.itb.ac.id)
[2]Department of Physics, Universitas Negeri Padang, Padang, Indonesia
(Tel : +62-813-2102-9889; E-mail: ramlisutan@ymail.com)
[3]Department of Physics, Institut Teknologi Bandung, Bandung, Indonesia
(Tel : +62-812-2111-4973; E-mail: dudung@fi.itb.ac.id)
[4]Department of Physics, Institut Teknologi Bandung, Bandung, Indonesia
(Tel : +62-812-212-7181; E-mail: krijal@fi.itb.ac.id)

**Abstract**-Giant magnetoresistance (GMR) material has great potential as next generation magnetic field sensing devices, have magnetic properties and high electrical potential to be developed into various applications such as: magnetic field sensor measurements, current measurements, linear and rotational position sensor, data storage, head recording, and non-volatile magnetic random access memory (MRAM). Today, the new GMR materials based on organic material obtained after allowing for Organic Magnetoresistance (OMAR) was found in OLEDs (organic light-emitting diodes). This organic material is used as a spacer layer in GMR devices with spin-valve structures. Traditionally, metals and semiconductors are used as a spacer layer in spin-valve. However, several factors such as spin scattering caused by large atoms of the spacer material and the interface scattering of ferromagnetic with a spacer, will limit the efficiency of spin-valve. In this paper, we describe a new GMR materials based on organic material that we have developed.

## I. INTRODUCTION

The discoveries of giant magnetoresistance (GMR) [1] and tunneling magnetoresistance (TMR) [2] in metallic spin valves (SVs) and magnetic tunnel junctions (MTJ) have already found widespread applications in magnetic recording and memory. The GMR material promises some important applications, among others as magnetic sensor. The GMR sensor has many attractive features, for example: reduction size, low-power consumption, low price as compared to other magnetic sensors and its electric and magnetic properties can be varied in very wide range [3].

GMR materials show benefit compare to other low-magnetic sensor technology such as high sensitivity, low cost, low power and small size (see Fig. 1). Because of these benefits compare to other magnetic materials, technically it is not very complicated and the technique is rather new in the world. Recently, we has been successfully to deposition of GMR thin film based on inorganic material with sandwich structure, and we have found that about 70 % MR value in NiCoFe/Cu/NiCoFe sandwich [4, 5]. In this paper, we would like to describe a new giant magnetoresistance based on organic material by Opposed Target Magnetron Sputtering (OTMS) method.

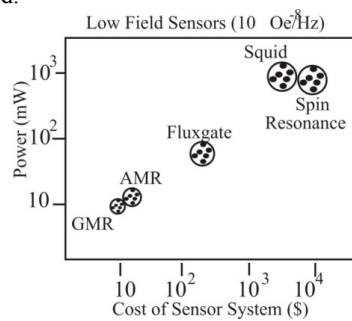

Fig. 1. Comparison of power, price and size of some magnetic sensors [3].

Organic magnetoresistance (OMAR) is a recently discovered, which magnetoresistance (MR) values up to 10% at room temperature and at fields of only a few millitesla have been reported in various organic materials [6]. Organic magnetoresistance effect also was found in organic thin films sandwiched between two conductive electrodes.

Currently, organic spintronics is a new and promising research field where organic materials are applied to mediate or control a spin-polarized signal. One of the most widely studied organic systems in this context is the spin valve consisting of an organic layer sandwiched between two ferromagnetic electrodes. Fig 2 shows the typical configuration of organic spin valve device.

A spin valve is a layered structure of two ferromagnetic (FM) electrodes separated by a nonmagnetic spacer. The spacer decouples the two FM electrodes and allows spin polarized carriers to travel through it without much relaxation. Traditionally, metals and inorganic semiconductors are used as the spacer material in SVs. However, a few factors like the spin scattering due to large atomic radii of the spacer materials and scattering from the FM/spacer interface, limit the efficiency of such SVs to a certain level. One solution to the first problem could be the introduction of spacer materials made from lighter elements, e.g. carbon, that will also have spin-transporting capabilities. The weak spin–orbit and hyperfine interactions of organic semiconductor small molecules and π-conjugated polymers lead to the possibility of preserving spin–coherence over times and distances much longer than the conventional metals or semiconductors. These which are the success of π-conjugated polymers and organic materials for commercial electronic applications, they were considered a promising alternative for spin transport [7].

The advantages of organic materials include chemical tuning of electronic functionality, easy structural modifications, ability of self-assembly and mechanical flexibility. These characteristics are exploited for large-area and low-cost electronic applications. For spintronics applications, organic spin valves are particularly attractive because of its flexibility and low cost.

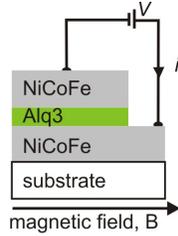

Fig. 2. Schematic picture of an organic SV device and magnetoresistance experiment.

## II. Experimental Method

Growth of spin valve of NiCoFe/Alq3/NiCoFe (see Fig 2) thin film performed at the Laboratory for Electronic Material Physics, Department of Physics, Institut Teknologi Bandung using dc-Opposed Target Magnetron Sputtering (dc-OTMS) method. Scheme of reactor dc-OTMS shown in Fig. 3. Sputtering target is NiCoFe as ferromagnetic material and Alq3 {(tris-(8-hydroxyquinoline) aluminum} as an organic material. Making the target NiCoFe performed by solids reaction with a molar ratio Ni:Co:Fe = 60:30:10. Alq3 target also making by solid reaction of Alq3 powder. The NiCoFe/Alq3/NiCoFe thin film was grown onto Si (100) substrate.

Samples of the NiCoFe/Alq3/NiCoFe were deposited in several different time of growth, so that they had different thickness of layers. Other deposition parameters are fixed. These parameters are: flow rate of Argon gas is 100 sccm, the growth pressure is 0.52 torr, dc Voltage is 600 volt, and the temperature is $100^0$C. The samples were characterized by using SEM (Scanning Electron Microscope) type JEOL JSM-6360 LA and magnetoresistance measurements were made by using a linear four-point probe method with current-perpendicular to-plane.

## III. Result and Discussion

Giant magnetoresistance ratio can be calculated by the equation MR (%) = {($R_{maks}$ - $R_{min}$) / $R_{maks}$} x 100%. The GMR ratio of NiCoFe/Alq3/NiCoFe thin film are shown in Fig. 4.

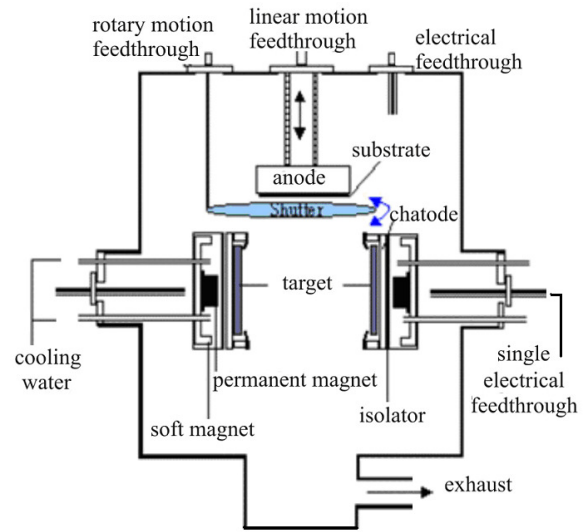

Fig. 3. Schematic picture of dc-OTMS reactor.

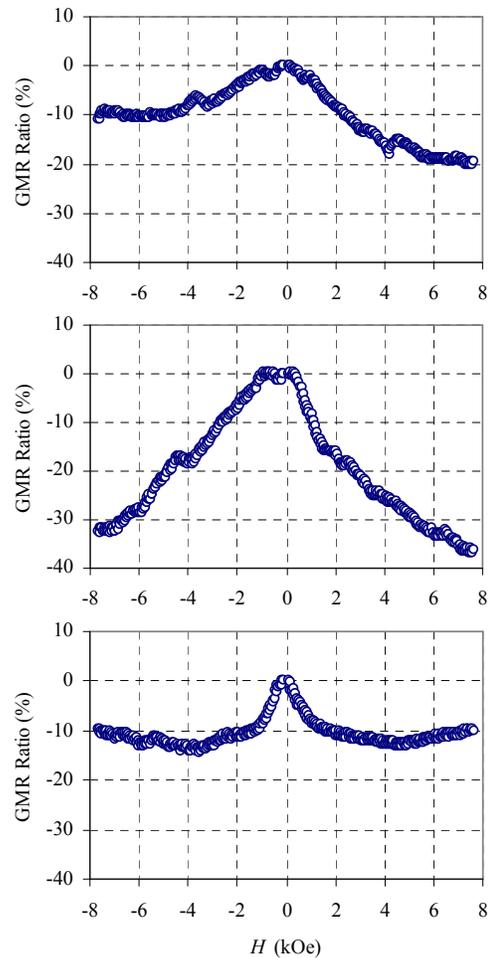

Fig. 4. The GMR ratio curve of NiCoFe/Alq3/NiCoFe at room temperature with time of growth: 10 minutes (top), 15 minutes (middle), and 20 minutes (bottom).

For the growth time of 10, 15 and 20 minutes, we found the GMR ratio respectively; 10.5%, 35.0% and 12.0%. The thickness of each layer for growth time of 10, 15, and 20 minutes are; NiCoFe (100 nm) / Alq3 (48 nm) / NiCoFe (100 nm); NiCoFe (137 nm) / Alq3 (72 nm) / NiCoFe (137 nm); NiCoFe (175 nm) / Alq3 (96 nm) / NiCoFe (175 nm), respectively. The maximum GMR ratio of about 35.5% at room temperature obtained at 15 minutes old growth.

As seen in the Fig.4, the GMR ratio is not monotonic decreasing to the thickness of the each layers of NiCoFe/Alq3/NiCoFe nor the spacer thickness as reported in [8]. It can be caused by variation of applied voltage in the layers growing process. In this experiment, for growth time 10, 15 and 20 minutes the average applied voltage are 78.4, 154.1, and 95.2 mV, respectively. It seems that the different applied voltage in layers growing process accidentally enhances the GMR ratio, as it is used in electrical conditioning [9]. We have found that as the applied voltage increasing, the GMR ratio is also increasing as shown in Fig 5.

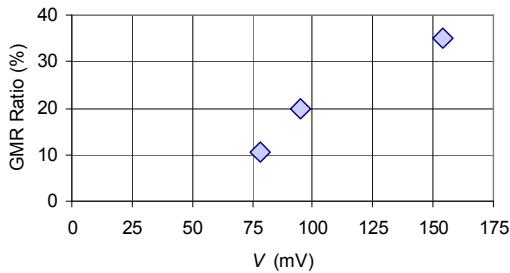

Fig. 5. Influence of applied voltage $V$ to GMR ratio.

It seems that the influence of spacer thickness is overridden by applied voltage that acts accidentally as electrical conditioning. Further investigation to merge both influences to GMR ratio is interesting but it is out of scope of this work.

The hysteresis loop of GMR spin valve is shown in Fig. 6. The shape of hysteresis loop shows a drop magnetization and indicates that there is a magnetic phases with the same coercivity.

## IV. CONCLUSION

We have been successfully grown a new GMR sensor materials based organic materials with a spin valve structure NiCoFe/Alq3/NiCoFe using Opposed Target Magnetron Sputtering (OTMS) onto Si (100) substrate. Magnetoresistance ratio of the sample are affected by time of growth. In this research, the maximum GMR ratio of 35.5% at room temperature obtained on time of growth is 15 minutes, where this maximum value of GMR ratio is caused mainly due to applied voltage about 154.1 mV during the layers growing process. From this results it was concluded that a sandwich NiCoFe/Alq3/NiCoFe that have been developed, has the potential to be used as magnetic field sensors.

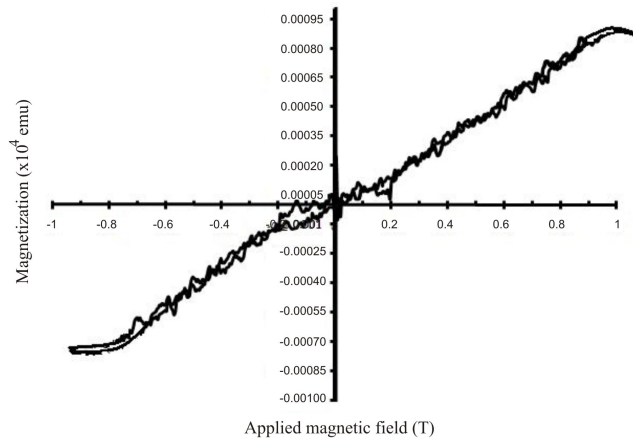

Fig. 6. Room-temperature hysteresis loop organic GMR spin valve, NiCoFe/Alq3/NiCoFe.

ACKNOWLEDGMENT

This work was (partially) supported by the Directorate for Research and Community Service, The Ministry of National Education Republic of Indonesia through Competency Grant under contract No: 407/SP2H/PP/ DP2M/VI/2010.